# Holistic structure of neural pathways underlies brain perceptual rivalry: Physical mechanism of auditory stream segregation

Yuxuan Wu[1], Jinling Gao[2], Xiaona Fang[3]* and Jin Wang[4]*

[1] Biophysics & Complex System Center, Center of Theoretical Physics, College of Physics, Jilin University, Changchun 130012, People's Republic of China

[2] State Key Laboratory of High Pressure and Superhard Materials, College of Physics, Jilin University, Changchun, 130012, China

[3] School of chemistry, Northeast Normal University, Changchun, Jilin, 130024, China

[4] Department of Chemistry and of Physics and Astronomy, Stony Brook University, Stony Brook, New York 11794, United States

& 1 & 2 contributed equally to this work and should be considered co-first authors.
*Corresponding author: Jin Wang, Xiaona Fang
Email: jin.wang.1@stonybrook.edu, Fangxn482@nenu.edu.cn

Brain perceptual rivalry, exemplified auditory stream segregation of competing tones (A_, B__, ABA_), serves as a core mechanism of brain perception formation. While increasingly recognized as determining by neural connections rather than specific neural groups, the neuromechanism of brain perception remains uncertain. We demonstrate that auditory stream segregation arises from the topological structure of holistic neural pathways. By constructing a holistic pathway model using existing neurophysiological data, combining nonlinear neural dynamics and nonequilibrium physics, we uncover the biophysical mechanism of perceptual phase transitions from integrated (ABA_) to segregated streams (A_ or B_), as well as the mechanism of temporal dynamics, perceptual switching path, and attention regulation underlying these transitions. Further, we demonstrate how our framework reveals energy consumption of the auditory system and combines it with neuroelectrophysiology. Two psycho-acoustic experiments validate our predictions of perception alternation and attention modulation. Our framework provides a transformative perspective on how brain networks generate complex perceptual experiences, emphasizing the significance of neural pathway structure in the process of brain function realization.

## INTRODUCTION

It is increasingly evidenced that brain function emerges from holistic, systematic interactions through complex neural connections [1-5]. Connectomics studies reinforce this perspective, revealing its relevance in perceptual behaviors [6]. For perception, binocular rivalry involves multiple visual pathways beyond the primary visual cortex V1 [7,8], while auditory stream processing shows progressive separation along the ascending auditory pathway [9]. However, it remains unclear whether the holistic structure of neural pathways itself contains the mechanism underlying these perceptions. With the continuous progress in graph theory, complex network dynamics, and connectomics in recent decades, answering this question is prospective to bridge between physics and neuroscience.

To address this question, we developed a biophysical model of auditory stream segregation, which serves as a representative case of both auditory scene analysis and perceptual rivalry. Auditory streams represent the perceptual organization of sound sequences wherein distinct components of an auditory scene are grouped into separate streams [10-12]. Research typically employs a paradigm in which listeners hear repeating ABA_ triplet tones [13-18]. As the frequency difference ($\Delta f$) between A and B tones increases, listeners experience three distinct perceptual phases: First, *Integration* phase, during which only integrated stream of continuous ABA_ sequence is perceived; second, *Rivalry* phase, known as "auditory bistability" or "ambiguous perception", characterized by alternation between integrated stream and segregated streams of A_A_A_ or B__B__ sequences; and third, *Segregation* phase, marked by exclusive alternation between A_A_A_ and B__B__ as foreground and background respectively [19].

Despite numerous experimental findings over the years, opinions vary on the underlying neural mechanism of the auditory stream segregation. Existing theoretical frameworks, including abstract neural networks [14,20], events and chain organization models [15,16], Gestalt psychology-based interpretations [21], and others [22-25], have advanced understanding through diverse methodological lenses. However, they predominantly emphasize abstract, phenomenological, or empirical frameworks, which limit the extent to which they engage with fundamental neurobiology. Approaching this issue from the perspective of holistic neural connectivity and incorporating recent neurophysiological advances, we systematically organize the complete auditory neural circuit, encompassing both ascending and descending pathways, to extract a topological structure for our neural circuit model. This model exhibits three key characteristics: First, adherence to established neurophysiological anatomy, avoiding abstract or phenomenological constructs; Second, comprehensive representation of the entire neural pathways, transcending region-specific analyses (e.g., cochlea or primary auditory cortex); And third, reliance solely on fundamental mechanisms including sigmoidal stimulus-feedback functions, neural adaptation, neural spontaneous activity and decay, and modulation of $\Delta f$ and attention.

The analysis of such a neural pathway network requires more than nonlinear dynamics alone. Attractors and their alternations cannot fully explain several critical phenomena: the differential stability between perceptions, the mechanism underlying perceptual durations and alternation frequencies, and the preference for specific perceptual switching paths, etc. These questions root in the natural nonequilibrium dynamics of the system. By identifying the interactions and noise among neural networks, we can quantify the probability weights ($P_{ss}$) or nonequilibrium potential landscape $U = -\ln(P_{ss})$ of all system states [26]. Each perception manifests as a basin, analogous to a potential well, separated by barriers on the landscape. Consequently, perceptual switching, adhering to noisy neural dynamics, is characterized by escape from one basin through barrier crossing, followed by attraction to another basin. This switching is determined by two driving forces: (1) The landscape gradient ($-D\nabla U$), primarily representing the deterministic driving force towards the current attractor; And (2) $\boldsymbol{J}_{ss}/P_{ss}$, a generalized curl force derived from steady-state probability flux $\boldsymbol{J}_{ss}$ that destabilizes the current attractor and facilitates potential perceptual switching. Accordingly, global average of $\boldsymbol{J}_{ss}$ quantifies the nonequilibrium driving force intensity.

Through the application of nonlinear and nonequilibrium dynamics, our model successfully replicates the three auditory phases as $\Delta f$ increases. The temporal dynamics of these phases, particularly the mean duration time (MDT) of auditory perceptions, is highly concerned experimentally [17,18]. This temporal dynamics exhibits several well-documented features, including specific MDT trends relative to $\Delta f$ or presentation rate, and the equi-duration location of integrated and segregated perceptions. Through combined stochastic simulation and barrier height quantification, our model accurately captures these MDT features and elucidates their underlying mechanisms.

The experimental manifestation of perceptual switching varies according to the most probable switching paths across the perceptual states. However, stochastic trajectory analysis alone proves insufficient for determining these paths, particularly in multi-state systems [27]. Stochastic path integral, detailed in the Materials and Methods section, enables precise identification of these dominant paths. This analysis reveals a crucial insight: regardless of whether $\Delta f$ is low or high, perceptual switching between A_A_A_ and B__B__ states occurs through ABA_ intermediation. Our experimental verification of this finding substantially validates the holistic structural model.

Our framework incorporates attention regulation within the auditory cortex module [28-30], establishing $\Delta f$ and attention as complementary dynamic parameters. Simulation results confirm previous experimental observations that attention shifts lead to changes in the relative weights of auditory perceptions [31,32]. Subsequently, our analysis of attention effects on the landscape provides an explanation for the mechanism of auditory scene analysis.

ATP hydrolysis powers several critical neural processes, including ion channel switching during action potential transmission, synaptic vesicle filling, and neurotransmitter release. The energetics of these processes has garnered increasing scientific attention [33]. Nonequilibrium thermodynamics describes these mesoscopic behaviors, wherein ATP hydrolysis provides thermodynamical driving for operation and function, while the system dissipates free energy through heat exchange [34]. This mesoscopic thermodynamical mechanism, quantified by entropy production rate (*EPR*) detailed in Material and Method section, maintains macroscopic auditory system operation under nonequilibrium steady state (NESS) as reflected by flux $\boldsymbol{J}_{ss}$ [35,36]. Thus, average flux $\boldsymbol{J}_{ss}$ ($avv\ \boldsymbol{J}_{ss}$) and *EPR* reveal the nonequilibrium dynamical activity (i.e. nonequilibrium driving force intensity) and energy cost of auditory neural pathways, providing a dynamical and thermodynamical perspective on the auditory stream segregation process. Our results demonstrate that higher dynamical activity leads to the more intensive switching among perceptions, consequently requiring greater free energy expenditure, while maximum values occur in the vicinity of equi-duration location. These characteristics manifest in the difference between forward and backward cross-correlations ($\Delta corr$), which represents time irreversibility or the degree of detailed balance breaking [37-39]. The $\Delta corr$ calculation from finite time series data provides a crucial link between theoretical framework and neuroelectrophysiological experiment, as such data can be obtained through various invasive or non-invasive brain experimental methods. Further dynamical and thermodynamical analysis in *Supplementary Information* reveals significant natures of the attention mechanism.

This study includes two psychoacoustical experiments. Experiment I validates the theoretical prediction that integrated perception serves as an intermediary in the switching between foreground and background segregated streams, supporting the validity of our holistic structural model of neural pathways. Experiment II confirms that varying degrees of listener distraction produce corresponding increases in the weight of integrated perception, substantiating our findings regarding the attention mechanism. These experimental results demonstrate the efficacy of a systematic, holistic, and physical approach in understanding both auditory stream segregation and brain perceptual rivalry.

# RESULTS

## Dynamical model of
## the entire auditory neural pathways and frequency-depended landscapes

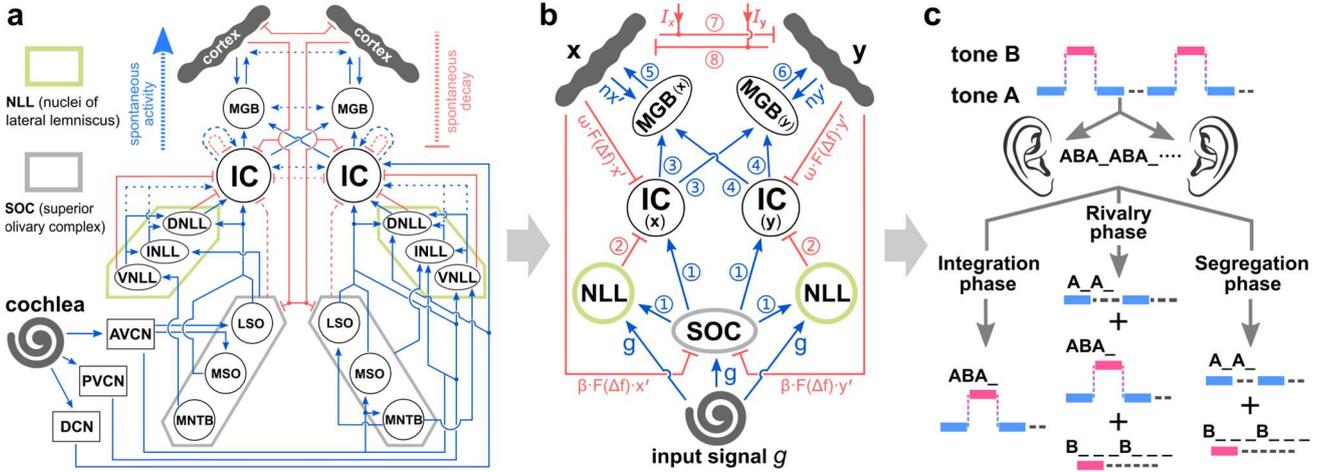

**FIG.1**. **a:** The auditory neural feedback pathways that may contribute to the auditory stream segregation and auditory scene analysis. Abbreviations of the pathway components are explained in main text. Only one side of the binaural pathways is displayed. Blue arrows represent the excitatory auditory neural pathways, while red bars represent the inhibitory ones. Dashed lines represent the pathways which are not ascertained or directly effective to the auditory stream segregation. Spontaneous neural activity and decay are represented as global characteristics. AVCN, PVCN, DCN: anteroventral, posteroventral, dorsal cochlear nucleus. LSO, MSO: lateral, medial superior olivary nucleus. MNTB: medial nucleus of the trapezoid body. DNLL, INLL, VNLL: dorsal, intermedial, ventral nucleus of lateral lemniscus. IC: inferior colliculus. MGB: medial geniculate body. **b:** The entire auditory neural circuit model after abstraction. The serial numbers and components beside pathways correspond to those in Eq.(1). **c:** Illustration for the auditory stream segregation.

The auditory neural feedback pathways involved in stream segregation comprise both ascending and descending components, being universal for mammals (Fig.1(**a**)) [40-43]. Neural projections in ascending auditory pathway are predominantly excitatory, progressively processing sound signals decoded by the cochlea and cochlear nuclei through hierarchical neural nuclei until reaching the auditory cortex. Frequency identification and stream formation occur incrementally along this pathway. Due to neurophysiological and anatomical similarities, LSO, MSO, MNTB and DNLL, INLL, VNLL are functionally integrated as SOC and NLL, respectively. Especially, NLL provides critical ascending inhibitory inputs to IC. Descending pathway, originating from the auditory cortex, primarily projects inhibitory signals to IC and SOC, modulating frequency selectivity, enhancing complex signal processing, and preventing gain dysregulation. Besides, the crucial spontaneous activity serves as a background signal and intrinsic noise in the neural system, enhancing the accuracy of analysis of sound signals [44,45]. Abundantly detailed interpretations and neurophysiological supports of the auditory neural pathway and following model design are provided in *Supplementary Information*.

Based on the neural pathway (Fig.1(**a**)), we derive a comprehensive model (Fig.1(**b**)), where $x$, $y$ model the firing rates of cortical neural populations associated with distinct iso-frequency bands on the auditory cortex. This originates from the tonotopic organization of both the cortex and subcortical nuclei [40,46,47]. We posit that $x$ and $y$ correspond to tones A and B respectively. The mutual excitations reflect that the cortex and the MGB perform as an entirety. Intracortical inhibitions between $x$, $y$ populations are modulated by attention coefficients $I_x$, $I_y$, and the function $F(\Delta f) = \ln(1 + \Delta f/v)$ based on the frequency difference $\Delta f$ of the input sound signal (No.7 and 8 in Eq.(1)). Logarithmic $F(\Delta f)$ models the nonlinear increase of mutual inhibition between iso-frequency bands. As $\Delta f$ increases, inter-band distance exceeds the range of strong intracortical connectivity to weaker broad-range interaction, resulting in a gradual saturation of inhibition strength. This frequency modulation $F(\Delta f)$ also influences corticofugal inhibitions to IC and SOC (No.1, 3 and 4 in Eq.(1)). Given that the key characteristics, $\Delta f$, and subsequently discussed tones duration are extracted into intracortical and corticofugal inhibitory projections by the pathway, we represent the input signal as a constant $g$.

We model the processing of input signals from lower hierarchies by each nucleus and its ascending output as elementary sigmoidal functions: the Logistic function $1/(1+\exp(-z))$ or Tanh function $1+\tanh(z)$, with the latter chosen for computational efficiency. Hence, functions $F$ for nuclei SOCs, ICs and MGBs, and mutual inhibitions $F_{cortex}^x$, $F_{cortex}^y$ read:

$$
\begin{aligned}
&① \quad F_{SOC} = \alpha[1 + \tanh(g - \beta F(\Delta f)(x' + y'))] \\
&② \quad F_{NLL} = \eta[1 + \tanh(g + F_{SOC})] \\
&③ \quad F_{IC}^x = \gamma[1 + \tanh(F_{SOC} - F_{NLL} - \omega F(\Delta f)x')] \\
&④ \quad F_{IC}^y = \gamma[1 + \tanh(F_{SOC} - F_{NLL} - \omega F(\Delta f)y')] \\
&⑤ \quad F_{MGB}^x = 1/\{1 + \exp[(F_{IC}^x + F_{IC}^y + nx' - \theta)/k]\} \\
&⑥ \quad F_{MGB}^y = 1/\{1 + \exp[(F_{IC}^x + F_{IC}^y + ny' - \theta)/k]\} \\
&⑦ \quad F_{cortex}^x = I_x F(\Delta f)x' \\
&⑧ \quad F_{cortex}^y = I_y F(\Delta f)y'
\end{aligned} \quad (1)
$$

where the serial number and components including input signal $g$, corticofugal inhibitory projections $\omega F(\Delta f)x'$, $\omega F(\Delta f)y'$, $\beta F(\Delta f)x'$, $\beta F(\Delta f)y'$, and Cortex-MGB mutual excitations $nx'$, $ny'$ are tagged in Fig.1(**b**), and

$$
\begin{aligned}
x' &= x - \theta_x \\
y' &= y - \theta_y
\end{aligned} \quad (2)
$$

with firing thresholds $\theta_x$ and $\theta_y$. Given the adiabatic processing of neural impulse conduction among nuclei and increased adaptation complexity at higher hierarchies [48], we incorporate adaptation functions $\Lambda(x) = a - b\tanh(x' - c)$ and $\Lambda(y) = a - b\tanh(y' - c)$ into the final auditory stream processing at cortex. Besides, the spontaneous decay $\rho x'$ and $\rho y'$ with strength coefficient $\rho$, and the spontaneous activity $\sigma$ as inherent background signal, are incorporated into the final expression. The Appendix provides coefficient values. Finally, the deterministic nonlinear dynamical equations are:

$$
\begin{aligned}
\frac{dx}{dt} &= F_x = \sigma - \rho x' + \Lambda(x) F_{MGB}^x - F_{cortex}^y \\
\frac{dy}{dt} &= F_y = \sigma - \rho y' + \Lambda(y) F_{MGB}^y - F_{cortex}^x
\end{aligned} \quad (3)
$$

Setting the baseline attention coefficient $I_x = I_y = 0.0285$ and increasing $\Delta f$ from 0 to 1 reveals the global dynamical evolution process of auditory stream segregation, which arise from the interaction among auditory pathway wiring, according to its neurophysiology: $\sigma$ enables fine-grained analysis of overall sound signals, while $-\rho x'$, $-\rho y'$ prevents over-excitation. Ascending pathway conveys the mixed excitatory signals of A and B tones to the cortex, while $\Delta f$-related intracortical and corticofugal inhibitions tune the signal processing (See section 1 and 2 in *Supplementary Information*). Initially, weak intracortical inhibition facilitates excitations of $x$ (for A tone) and $y$ (for B tone) merge into a monostable integrated state (*Int*) (Fig.2(**a**)). As $\Delta f$ increases, intracortical inhibition sharpens selective responses to A and B frequencies, leading to perceptual exclusivity, while corticofugal inhibitions preserve the concurrent sensitivity on both frequencies. Whereupon, the system undergoes the solitary bifurcation at $\Delta f \in (0.4218, 0.4219)$. At this bifurcation, three saddles partition the phase plane into four distinct regions: two integrated states (*Intx*, *Inty*) representing fusion of bilateral $x$ and $y$ excitations, corresponding to the integrated stream perception (ABA_ABA_), and two segregated states (*Segx*, *Segy*) respectively corresponding to the exclusive excitation of $x$ or $y$, representing segregated stream perceptions (A_A_ and B__B__) (Fig.2(**b**)). Further increases in $\Delta f$ intensify the polarity of *Segx*, *Segy*, meanwhile drifting *Intx*, *Inty* apart (Fig.2(**c**)). In this nonlinear dynamical framework, each perception type (segregated or integrated) comprises two states, because one tone (A or B) occupies the perceptual foreground while another recedes in background, although this distinction becomes negligible for integrated perceptions *Intx*, *Inty* [19,49].

Introducing noise into Eq.(2) transforms it into a stochastic differential equation (Eq.(4)), and the corresponding Fokker-Planck equation can be obtained (Eq.(5)). Consequently, we demonstrate the potential landscape $U$ and probability flux $\boldsymbol{J}_{ss}$ (Fig.2(**a-c**)). As discussed in Introduction, under NESS, $-D\boldsymbol{\nabla}U$ and $\boldsymbol{J}_{ss}/P_{ss}$ manifests the combined effect of $\boldsymbol{F}$ and noise. On the landscape, perceptual states $Int\ x, y$ and $Seg\ x, y$ form high-weight basins as "potential wells", with attractors virtually coinciding with basin minima. State switching requires overcoming $-D\boldsymbol{\nabla}U$ that points to current basin bottom. This switching driven by noise in stochastic dynamics can be described as landscape barrier crossing process in nonequilibrium dynamics. The diffusion coefficient $D$ modulates noise intensity: high $D$ yields low barriers and flat landscapes, while low $D$ leads to high barriers and rough landscapes (Fig.3(**e, f, g, h**)). Therefore, barrier height naturally quantifies the perceptual state stability, and correlates with auditory perception durations.

The nonequilibrium landscape reveals auditory dynamics as a phase transition process [17]. After dynamical bifurcation $\Delta f = 0.4219$, $Seg\ x$ and $Seg\ y$ basins emerge and deepen as $\Delta f$ increases, while at an imposed truncation $\Delta f = 0.8$, $Int\ x$ and $Int\ y$ basins lose stability as perceptual states (Fig.3(**h**)). Fig.3(**b-i**) illustrate perceptual weight evolution via landscapes, establishing $\Delta f = 0.4219$ and $\Delta f = 0.8$ as two practical phase transition points. On the other hand, due to noise sensitivity, biological systems function effectively only within effective $D$ range [50], as shown in Fig.3(**a**). For universality, we take the median of the $D$ range for each $\Delta f$ value as the optimal condition. The relationship between $D$ and presentation rate will be addressed later.

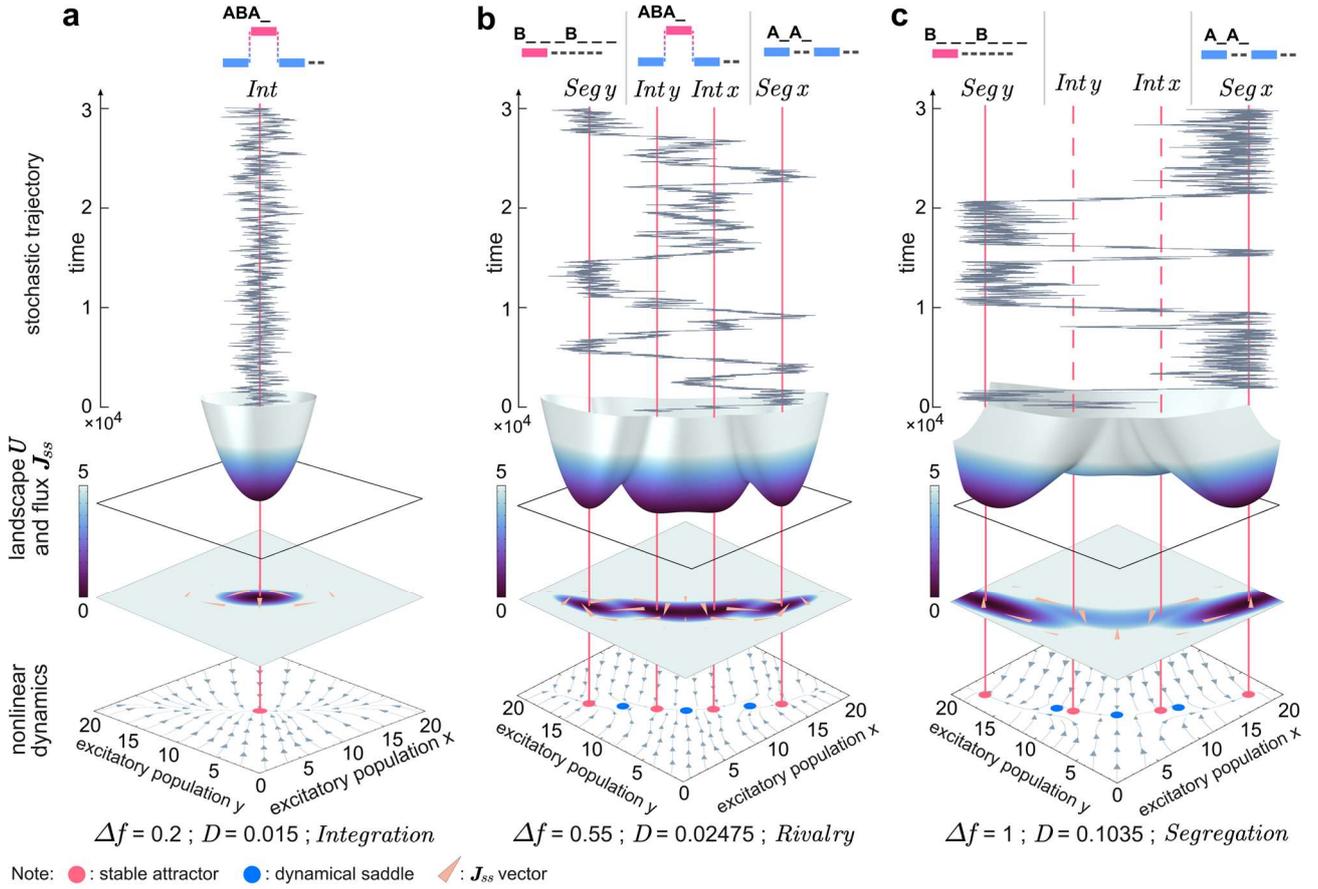

**FIG.2**. **a-c**: Illustrations of three phases throughout auditory stream segregation process, via nonlinear deterministic and stochastic dynamics, and nonequilibrium dynamics. In **a-c**, from bottom to top, there are streamline diagrams of deterministic nonlinear driving force $\boldsymbol{F}$ obtained by Eq.(3), two-dimensional vertical views of landscape $U$ with probability flux $\boldsymbol{J}_{ss}$, three-dimensional views of landscape $U$ obtained by Eq.(5), and 1-dimensional stochastic trajectories of switching between auditory perceptions obtained by Eq.(4). Magenta lines in **a-c**, which nail through the diagrams, connect the attractors of perceptions, corresponding landscape basins and stochastic trajectories. Dashed lines in **c** indicate that the weak stability of $Int\ x$ and $Int\ y$ can no longer form practical perception. For an explicit display, in each 2-dimensional landscape diagram, peach cones that depict the $\boldsymbol{J}_{ss}$ are taken the logarithm of the real $\boldsymbol{J}_{ss}$ values. They reflect the orientation of $\boldsymbol{J}_{ss}$, and where there are significant probability flows. Besides, for simulating the initial organization of auditory streams, each stochastic trajectory starts from $Int$ state (*Integration* phase) or saddle between $Int\ x$ and $Int\ y$ (*Rivalry* and *Segregation* phases). In **a-c**, $I_x = I_y = 0.0285$, and $D$ values are according to the optimal values (polygonal line in Fig.3(**a**)).

Combining nonlinear, stochastic, and nonequilibrium dynamics, our model replicates three phases in the auditory stream segregation phase transition process, as discussed in Introduction (Fig.2(**a-c**), Fig.3(**a**), Extended data Fig.1): *Integration* phase ($\Delta f \in [0, 0.4219)$), *Rivalry* phase ($\Delta f \in [0.4219, 0.8]$), and *Segregation* phase ($\Delta f \in [0.8, 1]$). As $\Delta f$ increases, the $Int$ basin representing integrated stream perception expands along the diagonal, forming a dynamical tristability with three basins, where $Int\ x$ and $Int\ y$ effectively function as a single basin ($Int$) due to the negligible barrier between them. Further increase in $\Delta f$ leads to the enhancement of $Seg\ x$ and $Seg\ y$ basins of segregated stream perceptions, accompanied by the decay of the $Int$ basin until it becomes equivalent to a thick barrier between *Seg x* and *Seg y*, rarely capturing states. Meanwhile, $\boldsymbol{J}_{ss}$ forms vortices at two barriers between *Int* and *Seg x*, *Seg y* basins, characterizing intensive perceptual switching under tristability. As $\Delta f$ increases, the two vortices become adjacent, representing the nonequilibrium dynamics of complete stream segregation. Two sub-basins *Int x* and *Int y* in basin of integrated stream relate to the initial auditory stream organization and foreground-background stream differentiation in integrated perception (See section 3 in *Supplementary Information*).

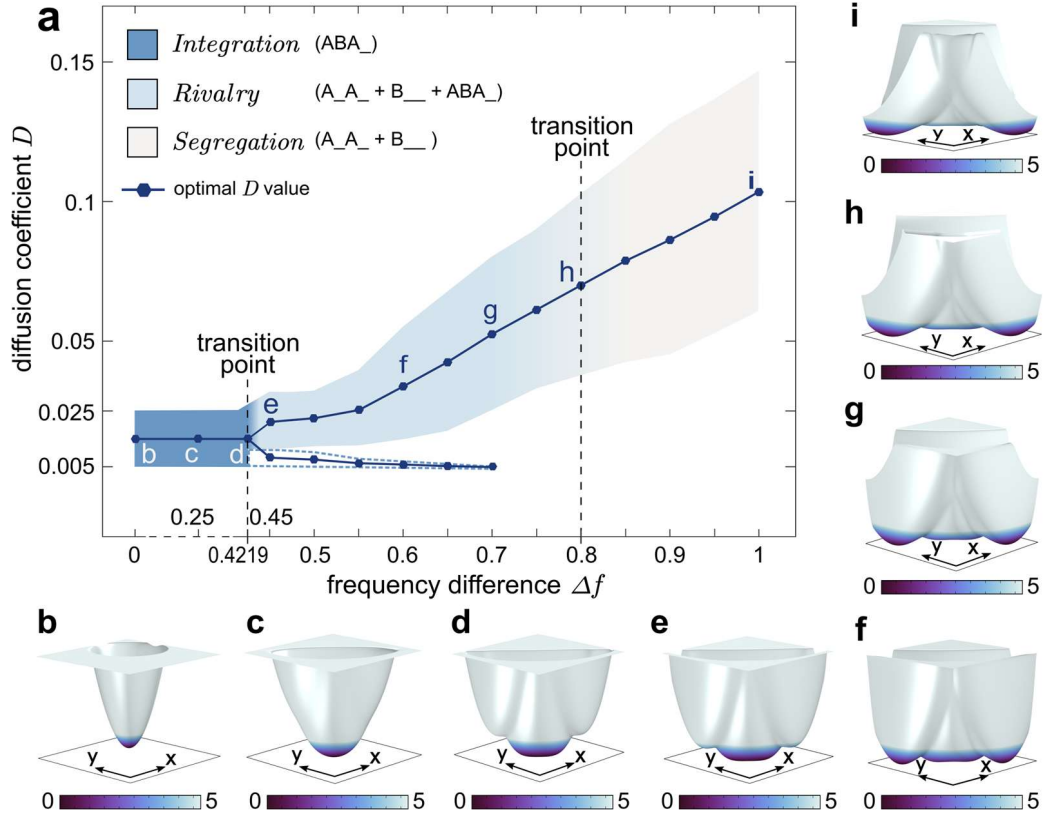

**FIG.3. a:** Effective $D$ range. *Integration*, *Rivalry* and *Segregation* respectively correspond to three auditory perceptual phases interpreted in the main article. The gradient color range $\Delta f \in [0.65, 0.8]$ is the transition region from the auditory bistability (dynamical tristability) to the auditory monostability (dynamical bistability) of segregated streams only. For each $\Delta f$ value, point on the polygonal line locates on the midpoint between upper and lower boundaries of the colored area. The area circled by dashed line is discussed in *Supplementary Information* section 3. **b-i:** 3D views of landscapes. Their positions mapped onto effective $D$ range are marked in **a**.

# Temporal dynamics:
# mean duration time and barrier height of the auditory perception

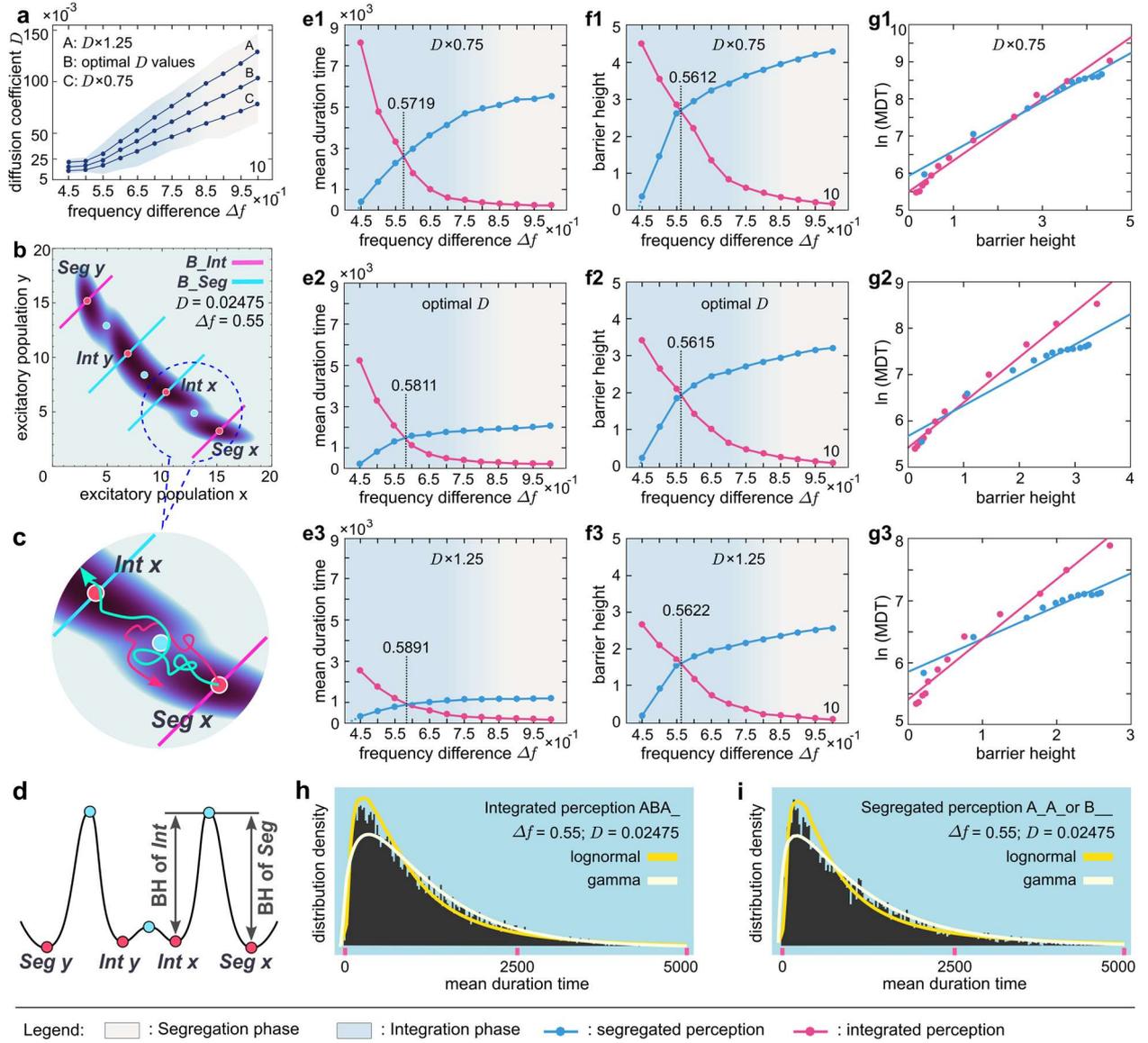

**FIG.4. a**: Portion of Fig.3(**a**). Polygonal line B of optimal $D$ values is the original line in Fig.3(a). $D$ values of points on line A are ones on B multiplied by 1.25, while values on C are ones on B multiplied by 0.75. **b**: Illustration of the method to measure the duration, using the case $\Delta f = 0.55$ as an instance. Magenta solid circles represent the stable attractors, and blue ones represent saddles. Two pairs of boundaries *B_Int* and *B_Seg* are parallel with diagonal from (0, 0) to (20, 20). **c**: Illustration of the stochastic trajectories starting from *Seg x*. States that barely cross the barrier possess a high probability to return back (magenta stochastic trajectory). They need to be close enough to the attractor to be captured by the basin (cyan trajectories). **d**: Definitions of BH of integrated (*Int*) and segregated (*Seg*) perceptions. (**e1-e3**, **f1-f3**): MDTs, and corresponding BHs of the side from *Int x (y)* to *Seg x (y)* (magenta lines), and of the side from *Seg x (y)* to *Int x (y)* (blue lines). In the calculation, the $D$ value of each sampling point is from the corresponding point in **a**. The $\Delta f$ values where BHs (MDTs) on both sides is equal are marked. (**g1-g3**): Linear fittings between ln(MDTs) and BHs. (**h**, **i**): Fitting of the distributions of the durations, selecting case $\Delta f = 0.55$ as the instance. For explicit display, distribution tails of rare events are cut off.

The mean duration time (MDT) is defined as the average duration time in the basin of attraction (quantified from 10,000 stochastic simulations in this study). A duration time represents the period a state takes to reach the boundary *B_Int* or *B_Seg* (Fig.4(**b**)), where for integrated perception, it spans the interval to reach either *B_Int* boundary from the saddle between $Int\ x$ or $Int\ y$ attractors, and for segregated perceptions, it tracks the period to reach the nearest *B_Seg* boundaries from $Seg\ x$ or $Seg\ y$ attractors. The boundaries ensure no relaxing and backtracking around the barrier region, indicating a complete perceptual transformation where the previous perception ceases functioning (Fig.4(**c**)). Correspondingly, barrier height (BH) is defined as shown in Fig.4(**d**). For integrated perception, this BH definition is practical, due to that states will rapidly move to attractor $Int\ x$ or $Int\ y$ from the saddle between them. Fig.4(**e1-e3**, **f1-f3**) exhibit the MDT and BH results under three distinct conditions from Fig.4(**a**).

We first reveal the fundamental MDT characteristics: Throughout *Rivalry* and *Segregation* phases, MDTs of integrated or segregated perceptions rapidly decrease or increase with $\Delta f$, intersecting at equi-duration locations where MDTs of both perceptions are equal, then softly tending to the minimum or maximum values respectively (Fig. 4(**e1-e3**)) [14,15]. The essence of MDT tendencies lies in that along with $\Delta f$, varying landscapes lead to varying perceptual stabilities. These stabilities are quantified by BHs (Fig.4(**f1-f3**)). In fact, given $U = -\ln P_{ss}$ (Eq.(7)), an exponential relationship between MDT and BHs persists, despite landscape deformations caused by $\Delta f$ variation (Fig.4(**g1-g3**)) [51]. Furthermore, the points with balance BHs on both sides near $\Delta f = 0.55$ unveil that the MDT equi-duration locaions result from the equal weights of landscape basins of integrated and segregated perceptions. The $\Delta f$ values of equi-duration locations are slightly larger than points with balance BHs, because that a wider basin of integrated perception leads to longer state relaxation times (Fig.2(**b**)). The same reason also explains why the maximum MDT of integrated perception at $\Delta f = 0.45$ exceeds that of segregated perception at $\Delta f = 1$ [15], despite similar barrier height differences.

MDT results throughout Fig.4(**e1-e3**) emanate a positive correlation between noise intensity $D$ and tone durations (also presentation rate, PR) (See *Supplementary Information* for sufficient explanation). Results of our experimental I in the following also support this. Its neurophysiological basis lies in that auditory phase locking, random spontaneous firing and short-term synaptic plasticity [41,46]. In Fig.4(**c**), the $D$ conditions, A, B and C for Fig.4((**e3**, **f3**), (**e2**, **f2**), and (**e1**, **f1**)), correspond broadly to 200ms, 150ms, and 100ms tone durations (PR = 5Hz, 6.67Hz, and 10Hz) [14,15]. In our model, $\Delta f = [0.45, 0.1]$ range maps to 3-20 semitones, with the equi-duration location near $\Delta f = 0.55$ corresponding to 5-6 semitones. Thus, our model is linked to experiments via temporal dynamics.

At low $\Delta f$ of 2-4 semitones (approximately $\Delta f$ of 0.4219-0.5 in our model), decreasing presentation rate increases the "difficulty" of perceiving integrated perception [52]. This occurs because MDT differences between integrated and segregated perceptions rapidly decline in $\Delta f \in [0.4219, 0.5]$ (Fig.4(**e1-e3**)). Due to the positive correlation between MDT and proportion [14], that increasing "difficulty" results from the decaying proportion of integrated perception. Similar reasons also apply to Ref.[53], which reports that as $\Delta f$ increases, the proportion of segregated perception raises faster when presentation rate is higher. Further, the locations of balance BHs gradually shift rightward on the $\Delta f$ axis as presentation rate decreases (Fig.4(**f1-f3**)). As the result, equi-duration locations in Fig.4(**e1-e3**) present the same behavior, as reported by Ref.[15]. These temporal dynamical behaviors are based on the evolution of nonequilibrium landscapes, which emerges from interactions within neural pathways.

Gamma and lognormal distribution of the perceptual durations have been studying for years [14,15,18,54-56]. Our results in Fig.4(**h** ,**i**) conform to both these distributions, with the lognormal distribution fitting better for the height of the distribution peak [14]. This entrenches the rationality of our temporal dynamical results.

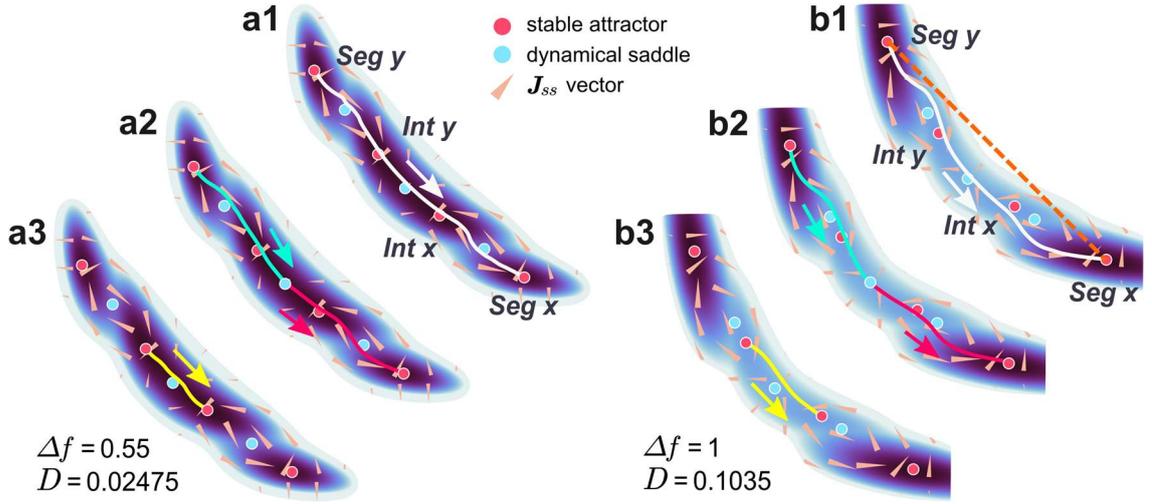

**FIG.5**. **a1-a3:** Dominant paths of the auditory bistability (dynamical tristability), *Rivalry* phase. **b1-b3:** Dominant paths of the pure segregated streams, *Segregation* phase. The dashed line in **b1** is a hypothetical switching path from $Seg\ y$ to $Seg\ x$ directly. For the convenient presentation, **a1-a3**, **b1-b3** are the landscape basin parts cropped from the 2-dimensional view of landscapes in Fig.2(**b**, **c**) respectively, and only the paths along the direction of corresponding arrows are shown (Ones of the opposite direction can be simply derived by symmetry). The magenta and baby blue solid circles represent the deterministic stable states and saddles respectively, as shown in Fig.2(**b**, **c**). In **a1** and **b1**, white lines are dominant paths from $Seg\ y$ to $Seg\ x$. In **a2** and **b2**, cyan lines are paths from $Seg\ y$ to the middle saddle between $Int\ x$ and $Int\ y$, while purplish-red lines are paths from the middle saddle to $Seg\ x$. In **a3** and **b3**, yellow lines are paths from $Int\ y$ to $Int\ x$. Note that as a rational approach, the middle saddles are accounted start and finish of the paths for the integrated states, according to the fact that $Int\ x$, $Int\ y$ perform as one state in these case.

The stochastic path integral method selects the most probable dominant path from multiple stochastic trajectories. Fig.5 visualizes these dominant switching paths for *Rivalry* and *Segregation* phases at $\Delta f = 0.55$ and $\Delta f = 1$ as examples. As queried in Introduction, an assumed direct path (orange dashed line in Fig.5(**b1**)) between $Seg\ x$ (A_A_A_) and $Seg\ y$ (B__B__) differs significantly from the actual dominant switching path. Whereas the assumed path appears shorter that conforms to traditional auditory stream segregation models, the real path involves an intermediate integrated perception (ABA_ABA_).

The dominant path eliminates the uncertainty above (Fig.5(**a1**, **b1**)), revealing a counterintuitive switching mechanism: Foreground-background perception transitions are mediated through integrated perception, especially for the Segregation phase. Our experiment I interpreted later confirms this prediction, demonstrating the intermediated switching, though the transient ABA_ perception is relatively difficult to be detected by subjects at large $\Delta f$ conditions.

Dominant paths and their converse, as well as the entire landscape, exhibits mirror symmetry about the diagonal from (0,0) to (20, 20). This signifies equal switching probabilities from ABA_ABA_ perception to either A_A_ or B__B__. Neither A tone nor B tone inherently holds a special status, although previous research suggests a potential propensity towards segregated stream of the higher tone [57]. We will revisit this nuance later via attention effect.

Dominant paths distinctly differentiate nonequilibrium dynamics from deterministic nonlinear dynamics. Unlike traditional attractor models, these paths do not pass through attractors or saddles, and do not coincide with their reverse paths. Even, the paths connecting *Seg x* and *Seg y* in Fig.5(**a1**, **b1**) diverge from the path combinations in Fig.5(**a2**, **b2**). All these characteristics reveal the irreversibility of nonequilibrium systems based on the interplay between $-D\nabla U$ and $\bm{J}_{ss}/P_{ss}$.

## Attention regulation in auditory stream segregation

Attention, represented by $I_x$ and $I_y$ for tones A and B (Fig.1(**a**, **b**)), is a dynamical parameter in auditory stream segregation that subjects can consciously regulate. Experimentally, attention tilts perceptual weight towards the focused perception. Our central finding of attention-modulation dynamics is (Fig.6(a)):

*Principle*: For auditory dynamical phases (e.g., Fig.2(**a-c**)) mapping onto $\Delta f$ range $[0, 1]$ at baseline attention, simultaneously amplifying or abating $I_x$ and $I_y$ will shrink or extend the $\Delta f$ range to $[0, \kappa < 1]$ or $[0, \kappa > 1]$.

Simulations in Fig.6(**b1-b4**, **b5-b8**) support the *Principle*. Particularly, landscapes with a fixed $\Delta f = 0.55$ in Fig.6(**b5-b8**) mirror those in Fig.3(**b**, **d**, **h**, **i**), demonstrating $\Delta f$ and attention as complementary dynamical parameters [58]. Specifically, increasing or decreasing $\Delta f$ or $I_x$, $I_y$ ($I_x = I_y$) will result in the same effect experimentally, while decreasing one can be compensated by raising another. It indicates that at extreme conditions of $\Delta f \to 0$ (almost identical A, B tones), forming segregated perceptions requires near-infinite attention on independent A and B tones. For instance, 20 times baseline attention is needed to induce a Rivalry phase at $\Delta f = 0.01$. Conversely, low attention makes segregated perceptions difficult to be formed within normal $\Delta f$ ranges. This frequency resolution limit experimentally exceeds the prediction based on human critical bands (10-15% of signal frequency) and reaches 3% [40]. Determined by our model, it stems from holistic pathway wiring including intracortical mechanisms beyond cochlear mechanics: Attention originating from the high-level auditory cortex or other functional brain regions can counteract frequency regulation effect across the entire pathway.

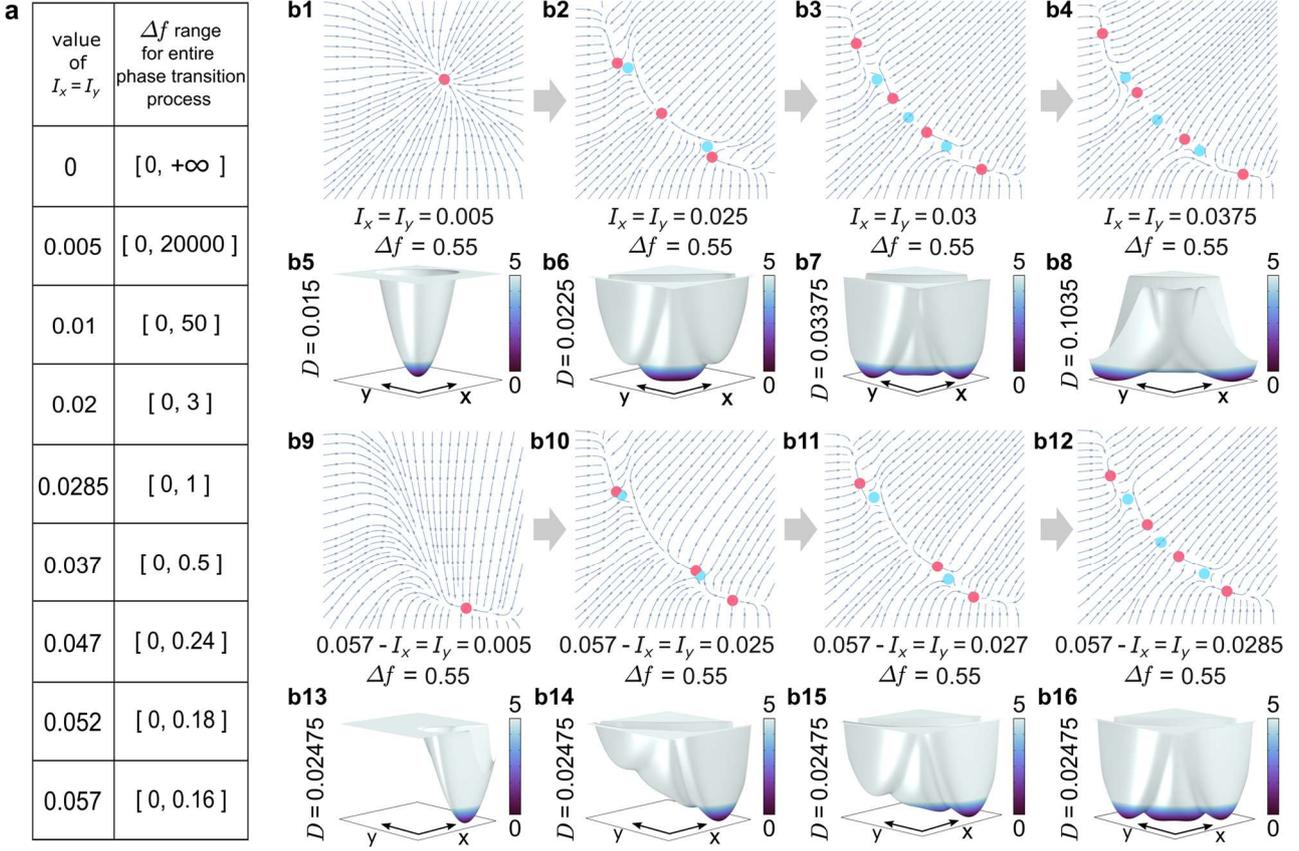

**FIG.6. a:** $\Delta f$ ranges for completing entire dynamical phase transition process under different values of $I_x = I_y$. **b1-b4, b5-b8:** Examples of deterministic dynamical structures of $\Delta f = 0.55$ with different values of $I_x = I_y$, and corresponding landscapes. $D$ values of these landscapes correspond to points b, d, h, i in Fig.3(**a**). **b9-b12, b13-b16:** Examples of deterministic dynamical structures of $\Delta f = 0.55$ with different values of $I_y = 0.057 - I_x$, and corresponding landscapes. $D$ values of these landscapes are selected properly in the practical range in Fig. 3(**a**).

While simultaneously increasing attention on both segregated perceptions (or A, B tones) is challenging for human listeners, simultaneous attention abatement is feasible. This enables us to test the *Principle* by regulating attention abatement levels. Except in the *Integration* phase, stronger attention abatement is expected to increase integrated perception dominance beyond that observed with baseline attention. Experiment II supports this anticipation.

More typically, listeners focus on specific perceptions, creating unequal attentions ($I_x \neq I_y$). From the *Principle*, dividing the phase plane diagonally from bottom-left to top-right, each side degenerates or evolves separately based on current $I_x$ and $I_y$. We presume $I_x + I_y$ remains constant at 0.057, due to that increased attention on one side naturally reduces attention on the other [29,59]. Fig.6(**b9-b16**) reveals that increased attention amplifies the weight of corresponding basin while diminishing or even eliminating the other, aligning with experimental observations (Fig.6(**b9, b13**)) [31,32]. Additionally, the narrow range of $I_x$ and $I_y$ for landscape deforming to monostability highlights the sensitivity of unequal attention modulation (Fig.6(**b10-b12, b14-b16**)) [60]. The above unequal attention mechanism unveils the underlying dynamics of auditory scene analysis: Attention selectively amplifies specific sound sources by deepening their corresponding basins on a multi-state landscape, while simultaneously shallowing other basins.

As shown by No.7 and 8 of Eq.(1) and Fig.1(**a**, **b**), this attention selectivity directly results from the intracortical unequal mutual inhibition (See section 2 in *Supplementary Information*).

In the dominant switching path section, we emphasized that some reports suggest higher switching probability from integrated perception to high-frequency segregated perceptions. This originates from natural attention imbalance [61]. During preparation for Experiment II, we observed most subjects more readily attracted to higher tone, revealing inherent deviations from balanced attention across individual listeners.

# Nonequilibrium dynamics and thermodynamics: driving force intensity, free energy cost and detailed balance breaking
# of auditory stream segregation

Here, we elucidate how our model reveals the nonequilibrium driving force ($\bm{J}_{ss}/P_{ss}$) intensity by average probability flux ($ave\ \bm{J}_{ss}$, Eq.(10)) and the required free energy cost by entropy production rate (*EPR*, Eq.(11)) of the entire auditory neural pathway. Meanwhile, it is pointed out how they are linked to neurophysiological experiments through quantitative and detailed balance breaking ($\Delta corr$, Eq.(12)).

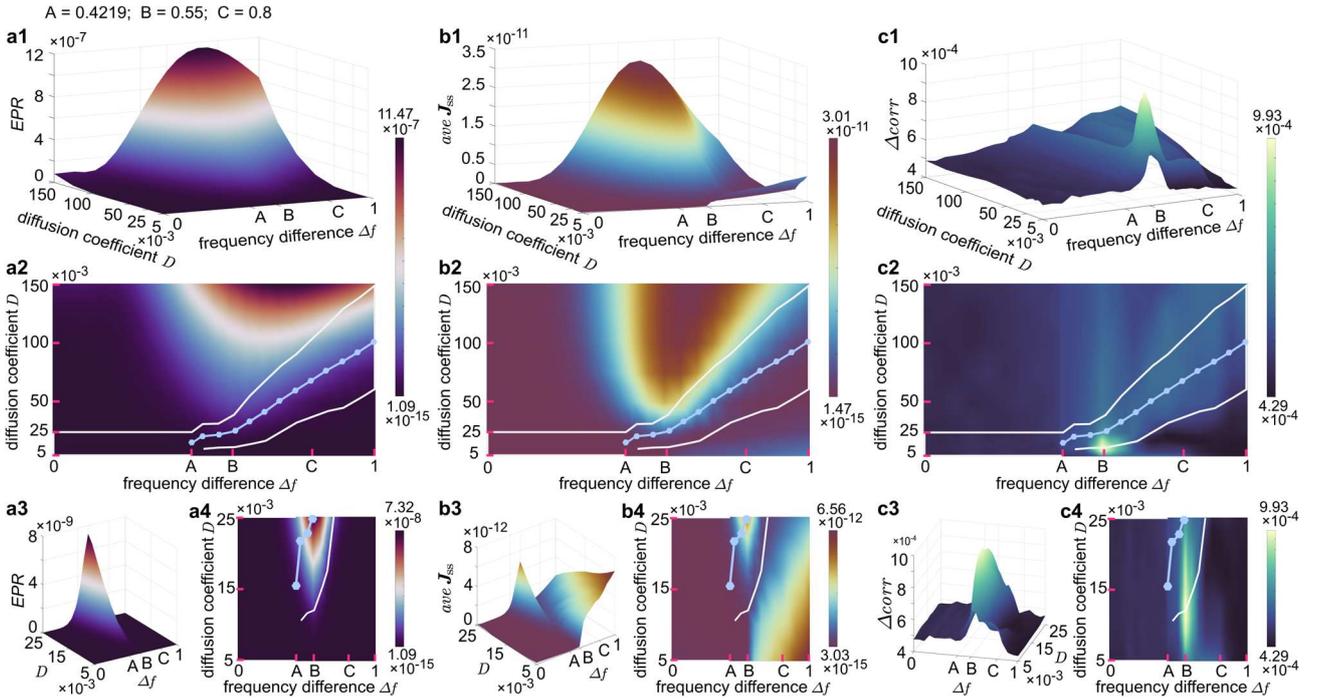

**FIG.7**. Illustrations of *EPR*, averaged $\bm{J}_{ss}$ and $\Delta corr$ in the coefficient space of $\Delta f$ and $D$. On the axis of $\Delta f$, A and C are bifurcations between *Integration*, *Rivalry* and *Segregation* phases, while B represents the case with three equally weighted basins. Subfigures **a2**, **b2**, and **c2** are the vertical views of the **a1**, **b1**, and **c1**, in which the areas with the white polygonal edges are the practical $D$ ranges according to Fig.3(**a**) (omitting area circled by dashed line, see *Supplementary Information* for more details), and the blue points on the blue polygonal lines are the optimal $D$ values after the bifurcation $\Delta f \in [0.4218,\ 0.4219]$. Subfigures (**a3**, **a4**), (**b3**, **b4**), and (**c3**, **c4**) show the situations of small $D$ values.

As discussed in the first section of Results, intensive nonequilibrium driving force quantified by probability flux $J_{ss}$ leads to frequent perception switching, i.e., high dynamical activity. Fig.7(**a1**, **a2**, **b1**, **b2**) reveal a precise parallel between $ave\ J_{ss}$ and *EPR*, determining corresponding requirement of free energy cost for supporting the nonequilibrium driving force. Within the effective $D$ range (Fig.3(**a**), Fig.7(**a2-c2**)), $ave\ J_{ss}$ remains low during the Integration phase until bifurcation $\Delta f \simeq 0.4219$ at the onset of the *Rivalry* phase and plateaus throughout the *Rivalry* and *Segregation* phases, despite a slight decline in the Segregation phase (Fig.7(**b2**)). The Rivalry phase exhibits the most frequent switching among three perceptions, creating the most intensive $J_{ss}$ across landscape basins (Fig.2(**b**)). In the *Segregation* phase, fewer switchings occur between segregated perception basins due to extended barriers formed by integrated perception (Fig.2(**c**)). Nonetheless, both *Rivalry* and *Segregation* phases show significantly higher dynamical activity than the *Integration* phase with only one basin. On the other hand, switching against landscape barriers requires free energy cost, i.e., work provided by ATP hydrolysis. *EPR* results in Fig.7(**a2**) demonstrate that both Rivalry and Segregation phases are sustained by high free energy consumption. It manifests that the auditory neural system expends more ATP to maintain multiple coexisting competitive perceptions compared to a single perception.

*EPR*, $ave\ J_{ss}$, and $\Delta corr$ exhibit remarkable peaks near $\Delta f = 0.55$ (MDT equi-dominant location), as shown in Fig.2(**b**), Fig.4(**d1-d3**, **e1-e3**), (Fig.7(**a2-a4**, **b2-b4**, **c2-c4**)). At optimal $D$ values (points on blue polygonal lines at $\Delta f = 0.55$), the interactions between switching frequencies across three equally weighted basins and barrier resistance achieve maximum efficiency. Consequently, both nonequilibrium driving force intensity and free energy cost reach their peak levels, so does $\Delta corr$ that reflects detailed balance breaking, also the degree of deviation from equilibrium.

$\Delta corr$, measuring time irreversibility and detailed balance breaking, reveals deviation from equilibrium. Deterministic nonlinear bifurcation $\Delta f \simeq 0.4219$ is reflected as a cliff-like step, while a continuous ridge in the practical $D$ range suggests a continuous phase transition (Fig.7(**c1**, **c2**)). Unlike the monostability of the *Integration* phase, *Rivalry* and *Segregation* phases display complex dynamical behavior with multiple attractive basins, which leads to a strong breakdown of the detailed balance of the system, namely intensely diverging from equilibrium. The correlation among $ave\ J_{ss}$, *EPR*, and $\Delta corr$ is not coincidental: detailed balance breaking is induced by directional probability flux $J_{ss}$, sustained by free energy cost and leading to time irreversibility. The most consistent features across these metrics arise near $\Delta f = 0.55$, where $\Delta corr$ peaks, reflecting its exceptional sensitivity (Fig.7(**a3**, **a4**, **b3**, **b4**, **c1**, **c3**, **c4**)).

A key advantage of $\Delta corr$ lies in characterizing *EPR* and $ave\ J_{ss}$ through their similar tendencies. It can be experimentally obtained by calculating the difference between forward and backward time trace cross correlations, requiring only neural population signal sequences. Such signals are widely measured in studies through fMRI [62,63], EEG or ECoG [64-67], and invasive cerebral electrophysiological methods [68,69]. We will revisit it in Discussion section.

## Psychoacoustic experiments

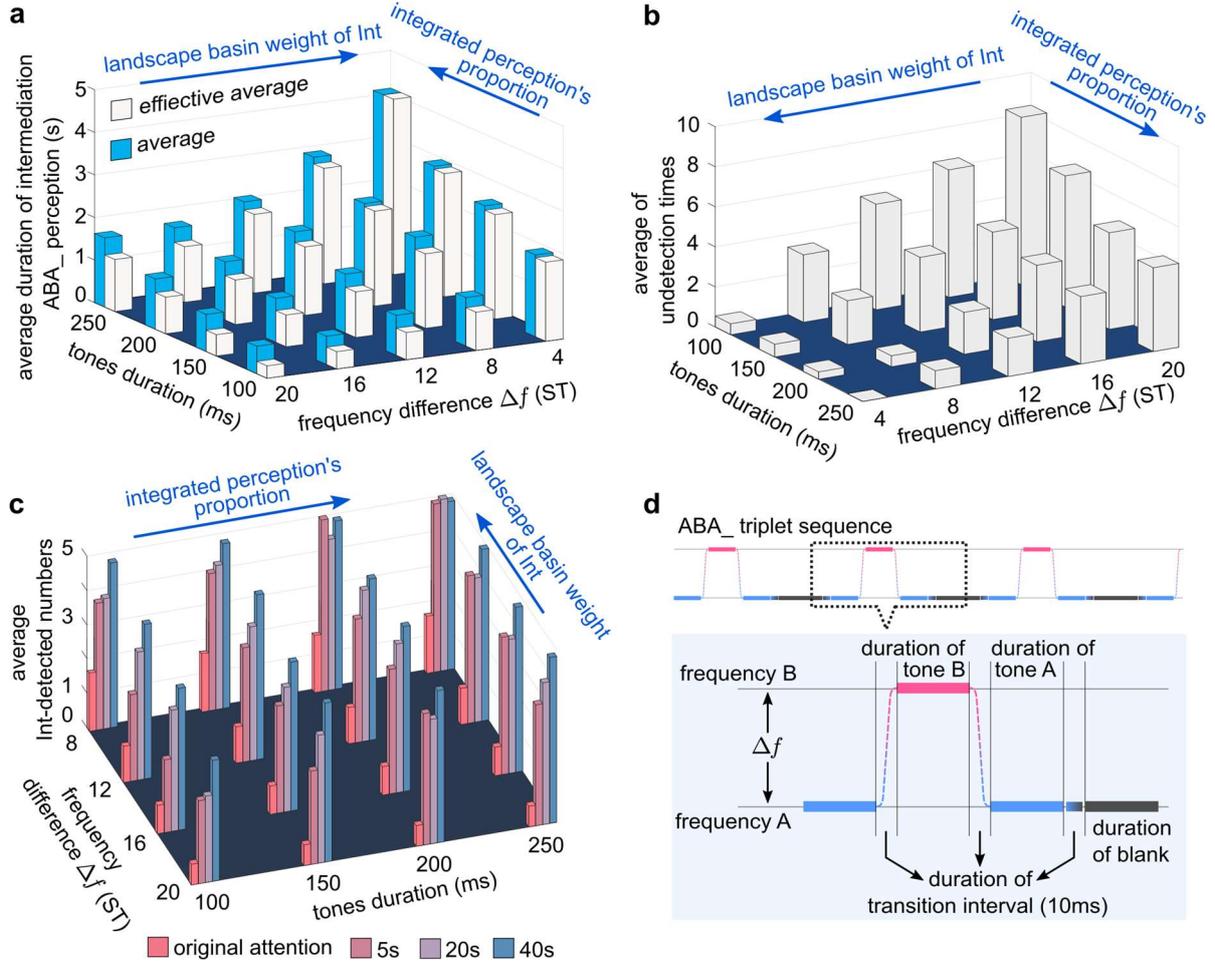

**FIG.8. a:** Average durations of the integrated perception for intermediation during the foreground-background switching of the segregated perceptions. "average" (blue bars) denote the results processed by Eq.(4), and "effective average" (white bars) denote the results processed by Eq.(5). **b:** Average numbers that the subjects did not detect the intermediating integrated perception, obtained by Eq.(6). **c:** Average numbers that the subject perceived the integrated perception after 5s, 20s and 40s distractions, and predicted results without distraction (original attention). According to results in temporal dynamics, basin weight change and proportion change of integrated perception ($Int$) are marked on **a**, **b**, and **c**. **d:** Illustration of the structure of the ABA_ auditory sequences.

Experiment I investigates whether the integrated auditory streams ABA_ABA_ ($Int$ perception) intermediates the foreground-background switch between segregated streams and A_A_A_ ($Seg\ x$ perception) and B__B__ ($Seg\ y$ perception) (Fig.8(**a**, **b**)). For each combination of $\Delta f$ and tones duration, across 15 subtests, we recorded the average duration of ABA_ perception detected by subjects during foreground-background transitions between A_A_ and B__B__ perceptions, and the average number that intermediary ABA_ was not perceived (Eq.(13-15), Methods). (Methods). Based on results of Fig.4(**e1-e3**, **f1-f3**), changes of basin weight and integrated perception ($Int$) proportion are noted as blue arrows. Assuming that switching is independent and random, the probability of paths A_A_ → ABA_ → B__B__ (and reverse path, white paths in Fig.5(**a1**, **b1**)) should be similar to direct A_A → B__B__ path (and reverse path, orange dashed line in Fig.5(**b1**)), with potentially higher probability for direct path. That is, column heights in Fig.8(**b**) should tend towards a uniform distribution. However, Fig.8(**b**) and Fig.8(**a**) coherently prove that increasing weight of *Int* basin or proportion of *Int* perception result in increasing detected frequency and duration of intermediary ABA_ perception, due to incremental attractiveness of *Int* basin. Both experimental results can only occur under the condition of A_A_ → ABA_ → B__B__ switching path, confirming that the switching between segregated perceptions is mediated by integrated perception. Besides, the barely detectable transient integrated perception, evident in stochastic trajectories (Fig.2(**b**, **c**)), contributes to the column height in Fig.8(**b**).

Fig.8(**a**) further validates our nonequilibrium dynamical framework. As $\Delta f$ decreases, relative stability of the integrated perception basin increases, resulting in longer mean duration of integrated perception (Fig.4(**f1-f3**)). Additionally, the results reaffirm the correlation between noise intensity and tone duration (or presentation rate) by showing synchronous increases in tone duration, integrated perception proportion, and average durations of intermediating ABA_ perception (Fig.4(**e1-e3**)). This dynamical structure arises from interactions across the holistic auditory pathway, as discussed previously.

Experiment II validates the *Principle* that attention diversion degenerates the dynamical structure to the condition of relatively low $\Delta f$, similar to decreasing $\Delta f$ itself. The number of detecting integrated ($Int$) perceptions after distraction served as the indicator, due to that decreasing $\Delta f$ results in increasing weight of $Int$ perception. Fig.8 (c) shows the average number of times that subjects detected $Int$ perception after experiencing 5, 20, or 40 seconds of distraction for each combination of $\Delta f$ and tone duration, while "original attention" columns are derived from our theoretical framework (See details in Methods). It reveals that columns for all distraction levels significantly exceed "original attention" columns, confirming substantial dynamical structure degradation. Longer distraction correlates with higher integrated perception probability. More specifically, as $\Delta f$ decreases, the results demonstrate that 5s distraction (2-5 average Int detections) degenerates the dynamical structure to the *Rivalry* phase, while 40s distraction (reaching 5 detections) approaches the *Integration* phase. These findings validate that weakened attention enhances the integrated perception basin when $\Delta f$ remains constant, which manifests degeneration of the dynamical structure. In other words, the $\Delta f$ range for the entire auditory phase transition process is elongated. Below-baseline attention involved in mutual inhibition between intracortical frequency bands weakens the stream segregation effect exerted by $\Delta f$ on the entire neural pathway. As predicted by the *Principle* in previous section of attention effect, attention and $\Delta f$ function as complementary dynamic parameters in auditory stream separation.

# DISCUSSION

**Universality of the holistic neural pathway model**: While our model is primarily based on neurophysiological data of specific species of mammals, Ref.[43] critically examines the fundamental neural network architecture, demonstrating structural and functional conservation of auditory neural pathways across mammalian species. The review highlights a conserved developmental blueprint that transcends species-specific variations, emphasizing the importance of overall neural network wiring. This neurophysiological and anatomical consistency substantiates the systematic organization of existing neurophysiological evidence, supporting the universal applicability of our model.

**Almost parameter-independent structural stability, and expansibility**. Parameter variations, by expanding or reducing values by 50%, reveals dynamical sensitivities. Modifying coefficients $k$ in $F_{MGB}$, $\gamma$ in $F_{IC}$, $v$ in $F(\Delta f)$, and $\sigma$, $\rho$ mildly altered the $\Delta f$ range for entire dynamical phase transition process (Eq.(1, 3)). Modifications to other parameters (including those in adaptation $\Lambda(x)$, $\Lambda(y)$ ) scarcely impact global dynamics.

Conversely, deleting or altering any connection in the neural pathway diagram or adaptation mechanism fundamentally disrupts global dynamics (Fig.1(**b**)). On the other hand, inherent scalability of our framework allows accommodation of more complex auditory scenes by parallel integration of additional cortical components (beyond $x$, $y$ ), each mapped to distinct tonotopically organized frequency bands. These findings underscore how the holistic topological structure of neural pathways fundamentally underpins brain perceptual functions.

**Intermediation by Integrated Perception**. Existing research suggests rapid perceptual switching potentially mediated by integrated perception, though often without discriminating foreground and background [14,18,32,70]. Ref.[57] explored auditory tristability using "energy landscapes," hinting at perception switching intermediation. Meanwhile, our dominant path analysis reveals that switching between A_A_ and B__B__ can be considered as two distinct probabilistic processes: A_A_ to ABA_ABA_, followed by a probabilistic switching within ABA_ABA_ to B__B__ (Fig.4(**a1**, **a2**, **b1**, **b2**)). This framework aligns with the hierarchical competition model of predictive coding [16], where the system temporarily selects the integrated organization before probabilistically switching to a segregated organization. These interpretations are compatible.

Rivalry between auditory perceptions, which is mediated by an integrated state, resembles a fundamental brain decision-making process. The dynamical behaviors exhibit remarkable similarities to the cognitive mechanism of "change of mind" in decision-making, particularly in attention regulation [71].

**Nonequilibrium dynamics, thermodynamics and Neurophysiology**. Despite growing theoretical interest in cerebral and auditory energetics [72-76], quantifying energy costs in large neural populations remains methodologically challenging. Our nonequilibrium dynamics and thermodynamics approach offers a promising solution. Non-invasive to invasive testing can collect $\Delta corr$ data without disrupting biological activity [62-69]. Critically, $\Delta corr$ and MDT results around $\Delta f = 0.55$ suggest a convergence between peak $\Delta corr$ and the MDT equi-duration locations, providing a crucial anchor across neurophysiology, thermodynamics, psychoacoustic data, and theoretical simulations.

Recent advancements in tonotopic cortical mapping enable tracking specific frequency positions ($x$, $y$ in our model), further enhancing interdisciplinary research [63,67,77]. Moreover, dynamical reconstruction of biological systems is emerging as a significant research frontier [78-80], with nonequilibrium dynamics and thermodynamic approaches increasingly capable of revealing phase transitions through time-sequence data analysis.

# METHODS
## Nonequilibrium physics methods

Eq.(3) represents an deterministic description about the firing rates of the cortical populations $x$ and $y$. However, a biological system inherently possesses internal and external randomness. We adopt the external white noise $\xi(t)$ with $\langle \xi(t)\xi(t')\rangle = \delta(t-t')$ to approach the intricate fluctuation of ion concentration, the sensitivity of ion channel, etc., among the nerve cells. Thus, Eq.(3) develops into the general Langevin equations:

$$\frac{dx}{dt} = F_1 + \sqrt{2D}\,\xi(t)$$
$$\frac{dy}{dt} = F_2 + \sqrt{2D}\,\xi(t) \tag{4}$$

where $D$ is the constant and isotropic diffusion coefficient as the matrix elements $D_{11} = D_{22}$ in diffusion matrix $\boldsymbol{D}$.

Eq.(4) reflect the stochastic dynamical behavior of the system. The Fokker-Planck equation can be derived from the stochastic differential equations Eq.(4):

$$\frac{\partial P(x,y,t)}{\partial t} = -\frac{\partial}{\partial x}(F_1 P) - \frac{\partial}{\partial y}(F_2 P) + D\left(\frac{\partial^2}{\partial x^2} + \frac{\partial^2}{\partial y^2}\right)P \tag{5}$$

Noting that the probability flux is

$$\boldsymbol{J} = \boldsymbol{F}P - D\boldsymbol{\nabla}P \tag{6}$$

Eq.(5) can be expressed in the form of $\partial P/\partial t = -\boldsymbol{\nabla}\cdot \boldsymbol{J}$. This is a probability conservation law where the change of the probability is equal to the net flux in or out. When the system reaches nonequilibrium steady state, it reads $0 = \boldsymbol{\nabla}\cdot \boldsymbol{J}_{ss}$, which indicates $\boldsymbol{J}_{ss}$ fluxes constitute the rotational field. The non-zero flux $\boldsymbol{J}_{ss}$ indicates that there is a net input or output flux. This breaks the detailed balance and leads the system to be in the intrinsic nonequilibrium state. Hence, the driving force for the stochastic dynamics can be decomposed into the flux $\boldsymbol{J}_{ss}$ and the landscape gradient therein as $\boldsymbol{F} = \boldsymbol{J}_{ss}/P_{ss} - D\boldsymbol{\nabla}U$, where

$$U = -\ln(P_{ss}) \tag{7}$$

is the generalized potential landscape. These compose the foundation of the nonequilibrium dynamics.

**Dominant path** possesses a maximum probability among all the connecting paths between arbitrary two states in the system. This is a derived result from the path integral of stochastic dynamics [81,82]. Resembling the well-known quantum path integral, it is critical to determine the effective Lagrangian, which is called Onsager-Machlup functional:

$$L_{OM} = \frac{1}{4D}[(\dot{x}-F_x)^2 + (\dot{y}-F_y)^2] + \frac{1}{2}\left(\frac{\partial F_x}{\partial x} + \frac{\partial F_y}{\partial y}\right) \tag{8}$$

Correspondingly, a certain path $\mathcal{X}$ with time span $[0,\tau]$ has the probability:

$$P[\mathcal{X}] = \exp\left(-\int_0^\tau L_{OM}\,dt\right) \tag{9}$$

Eq.(9), which belongs to two-points boundary value problem, is usually difficult to solve or seeking extreme value by Eular-Lagrange equation. However, based on the directly discretizing method of Onsager-Machlup functional [83-85], we employ genetic gradient annealing algorithm to seek the path that renders the integral in Eq.(9) to reach the minimal value.

**Average flux** $ave\ \boldsymbol{J}_{ss}$ represents the average level of the nonequilibrium driving force for dynamics. It can grasp the global dynamical activity of the system. in terms of that $\boldsymbol{J}_{ss}$ reflects the states flow on the landscape in a view of global probability transition. The definition of $ave\ \boldsymbol{J}_{ss}$ reads:

$$\int_\Omega P_{ss}|\boldsymbol{J}_{ss}|^2 d\Omega \quad \Omega\ is\ the\ domain\ of\ system \tag{10}$$

Given that $\boldsymbol{J}_{ss}$ features in the entropy production and the detailed balance breaking, the $ave\ \boldsymbol{J}_{ss}$ is expected a consistent tendency with them.

**Rates of entropy production and heat dissipation** reflect the properties of nonequilibrium thermodynamics and energetics. The general statistical physical entropy of the system is $S_{sys}=-k_B\int_\Omega P\ln(P)d\Omega$. Considering the Einstein's relation $D=k_BT$ and through some integral substitution, we have:

$$T\dot{S}_{sys}=\int_\Omega |\boldsymbol{J}|^2/P d\Omega - \int_\Omega \boldsymbol{F}\cdot \boldsymbol{J} d\Omega \tag{11}$$

$$=T\dot{S}_{tot}-T\dot{S}_{env}$$

where $T\dot{S}_{tot}$ is the total entropy production rate (*EPR*) of the system and environment, $T\dot{S}_{env}$ is the time change rate of heat exchange between the system and the environment, and $T\dot{S}_{sys}$ is the entropy production rate of the system itself [34,35]. In the crucial ATP hydrolysis process, the heat dissipation of each mesoscopic process can be converted to the work provided by hydrolysis to biomolecules, which is transformed into the system's free energy. When the system reaches steady state, $\dot{S}_{sys}=0$, and $EPR=T\dot{S}_{env}$. Therefore, although the system is usually non-Hamiltonian, we can still quantify the stochastic work, and dissipation cost of free energy.

**Time irreversibility and degree of the detailed balance breaking** originates from the asymmetric flow $\boldsymbol{J}_{ss}$ within the system. The difference between the forward and reverse cross-correlation:

$$\Delta corr = |\langle X(t)\cdot Y(t+\tau)\rangle - \langle Y(t)\cdot X(t+\tau)\rangle|$$

$$= \left|\int_\Omega d\Omega \cdot [X^i P^i(t) Y^j P^j(t+\tau) w^{ij}(\tau) - Y^i P^i(t) X^j P^j(t+\tau) w^{ji}(\tau)]\right| \tag{12}$$

$$\cong \left|\int_\Omega d\Omega \cdot XY\tau(J_X^{ij}+J_Y^{ij})\right|$$

is the effective statistical characteristic to measure the detailed balance breaking and time irreversibility, where $w^{ij}(\tau)$ is the transition probability of state $i$ switching to state $j$ during the interval $\tau$ [37,39]. It precisely quantifies the time asymmetry and non-zero flux between states among the entire system. The larger $\Delta corr$ is, the intenser the time irreversibility and detailed balance breaking are.

However, it is difficult to define $\tau$ and calculate Eq.(12) among the global continuous system. Practically, we use long time average to substitute ensemble average. That is, for each set of $D$ and $\Delta f$, the stochastic $x$, $y$ sequences of 300000 time span are generated, and we calculate the difference between the time forward and backward cross-correlations of $x$, $y$. Then, this procedure is repeated for 300 times and averaged. Finally, $\Delta corr$ for a set of $D$ and $\Delta f$ is gained. As stressed in the main content, simulation based on the finite sequence instead of the complicated integral has experimental practicality.

# Psychoacoustic experiment methods

Ten healthy-hearing participants (average age 25.3, including 3 females) were recruited for two psychoacoustic experiments after providing informed consent for the procedure and data usage. The experiments were conducted in a soundproof chamber, with all subjects demonstrating a clear understanding of the presented auditory signal types. Stimuli were generated using MATLAB with its Psychtoolbox extension. Both the experimental data and accompanying code are made available alongside the article.

**Experiment I**: 20 tests were conducted on each subject. Each test includes a group of 15 subtests of a same long ABA_ sequence, listened by a subject through earphones. Each ABA_ sequence in different tests varied in $\Delta f$ (4ST, 8ST, 12ST, 16ST, 20ST) and tones duration (100ms, 150ms, 200ms, 250ms, corresponding to 10Hz, 6.667Hz, 5Hz, 4Hz presentation rates). In one sequence, the durations of tone A, tone B and silent blank are equal, and included 10ms smooth frequency transitions (Fig.8(**d**)). During preparation, it is observed that high-frequency tone exhibits higher attraction for attention of subjects. Thus, tone A was configured to have doubled the repeating frequency of tone B to offset this perceptual bias.

After initial stream organization, if subjects perceived A_A_A_ as the more distinct foreground, they pressed " ← " on the keyboard. They were then allowed to slightly shift attention to B__B__ (foreground-background switch is not guaranteed to happen in every instance of segregated perception). Next, subjects pressed " ↓ " if they identified the integrated ABA_ABA_ sequence, then pressing " → " if they perceived subsequent B__B__. Time lengths between pressing " ↓ " and pressing " → " were recorded by the experiment program. If subjects directly switched to B__B__ without ABA_ intermediation, they pressed "0" (recording a zero time length). The impact on recorded intermediary ABA_ duration by the delay between identifying ABA_ABA_ (and pressing " ↓ ") and its actual occurrence in subjects perception was compensated by the next delay when identifying the following B__B__ and pressing " → ". Critically, the starting and ending perceptions of an effective measurement must be different segregated perceptions. For example, in the above process, if subjects perceived A_A_ again after detecting ABA_ and pressing " ↓ ", this measurement would be invalidated and excluded from the 15 subtests. The procedure remained identical when the subtest began with " → " for B__B__ perception.

Setting that for a certain couple of $\Delta f$ and tones duration, a data from the $i$-th subject ($i = 1, 2, ..., 10$) in the $j$-th subtest ($j = 1, 2, ..., 15$) is $X_i^j$, and the total number of the data being zero is $M_i$, the average duration of the intermediating $Int$ perceptions is defined as (blue columns in Fig.8(**a**)):

$$\frac{1}{10}\sum_i \frac{\sum_j X_i^j}{(15 - M_i)} \tag{13}$$

and the effective average duration reads (white columns in Fig.8(**a**)):

$$\frac{1}{10}\sum_i \frac{\sum_j X_i^j}{15} \tag{14}$$

The average number that the subjects did not detect the intermediating integrated perception is(columns in Fig.8(**b**)):

$$\frac{1}{10}\sum_i M_i \tag{15}$$

**Experiment II**: Auditory stimuli were identical to Experiment I (excluding $\Delta f = 4\text{ST}$ where integrated perception predominantly occurs). In each test combining specific tone duration and $\Delta f$, 15 subtests incorporated randomly sorted distraction time lengths of 5s, 20s, and 40s, with five repetitions per duration to quantify distraction levels. After establishing stable segregated streams (A_A_ or B__B__ perceptions), subjects pressed any keyboard key, then immediately diverted attention (e.g., silently checking a mobile phone). A transient warning signaled the end of distraction period, prompting subjects to refocus on the auditory stimuli. Subjects indicated integrated perception by pressing the "spacebar," and pressed any other key otherwise.

Two critical methodological considerations underpin the analysis. First, confirming the extent of dynamical structure degeneration required a nuanced approach. We leveraged MDT results (Fig.4(**a**, **e1-e3**)), using $\Delta f = 0.55$ and 1 as proxies for 5ST and 20ST, and $D \times 1.25$ and $D \times 0.75$ as proxies for 200ms and 100ms tone durations, respectively. Integrated perception proportions were calculated to represent detection numbers across 5 tests, establishing "original attention" columns in Fig.8(**c**). Second, dynamical structure degenerations under identical distraction conditions vary inherently among subjects. Within the *Rivalry* phase, integrated perception detection remains probabilistic. To mitigate this variability, results were statistically averaged across multiple subjects, enabling approximation of the true expected value within the altered dynamical structure.

**Appendix**

Here are the values of the coefficients in the model:

In $\Lambda(x)$ and $\Lambda(y)$, $a = 0.85$, $b = 0.15$, $c = 5$.

In $F(\Delta f)$, $v = 0.25$.

In Eq.(1), $g = 1$, $\alpha = 1.4$, $\beta = 0.028$, $\eta = 0.02$, $\gamma = 0.315$, $n = 0.25$, $\omega = 2$. For $F_{MGB}{}^x$ and $F_{MGB}{}^y$, $\theta = 0.0001$ and $k = 0.2$.

In Eq.(2), $\theta_x = \theta_y = 4$.

In Eq.(3), $\sigma = 0.25$, $\rho = 0.025$.

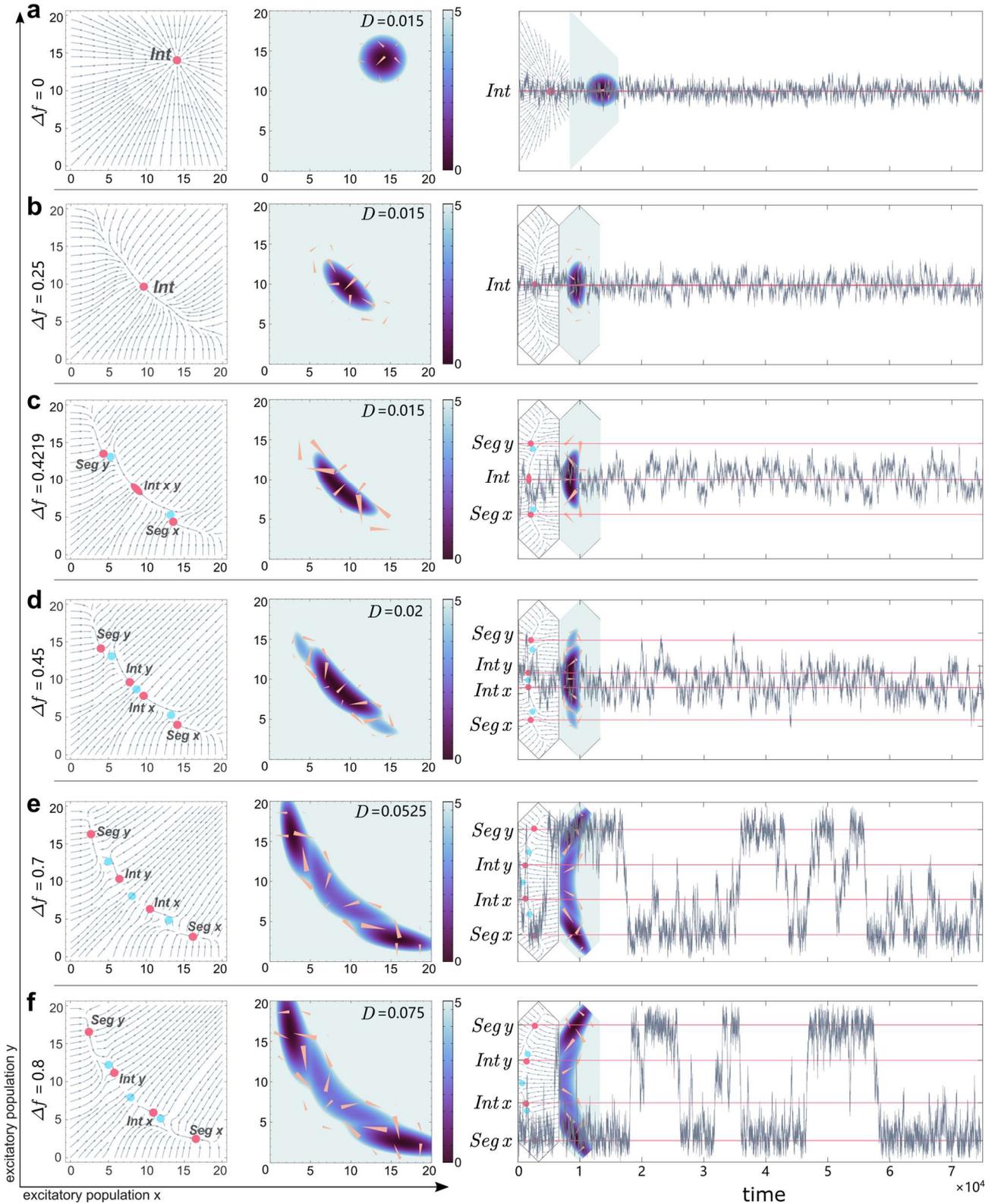

**Extended data FIG.1.** More detailed illustrations of different phases in the auditory stream segregation process. For each row of **a-f**, from left to right, these are the deterministic dynamical structure, the vertical 2-dimensional view of the landscape, and the 1-dimensional stochastic trajectory, which definitions are in accordance with those of Fig.2. The slices of dynamical structures and landscape 2-dimensional views are added into the subfigures of trajectories, for enhancing the connection between the deterministic nonlinear dynamics, nonequilibrium dynamics and the stochastic dynamics. Both the excitations of $x$, $y$ are very strong in **a**, due to that when $\Delta f = 0$, the excited iso-frequency bands on the auditory cortex are actually identical, without segregation and inhibitory regulations (See section 2 in *Supplementary Information* also).

## Acknowledgments

Yuxuan Wu thanks the beneficial discussion with Xiaochen Wang on attention mechanism and parameter dependency of the model.

Yuxuan Wu and Xiaona Fang thank the supports from National Natural Science Foundation of China (No.12234019 and No.32171245).